\documentclass[twocolumn,showpacs,aps,pre,floatfix,superscriptaddress]{revtex4}
\pdfoutput=1 
\usepackage{amsmath,amssymb,eucal}
\usepackage{graphicx}

\begin{document}
\title{Aggregation Driven by a Localized Source}
\author{P.~L.~Krapivsky}
\affiliation{Department of Physics, Boston University, Boston, MA 02215, USA}

\begin{abstract}
We study aggregation driven by a localized source of monomers. The densities become stationary and have algebraic tails far away from the source. We show that in a model with mass-independent reaction rates and diffusion coefficients, the density of monomers decays as $r^{-\beta(d)}$ in $d$ dimensions.  The decay exponent has irrational values in physically relevant dimensions: $\beta(3)=(\sqrt{17}+1)/2$ and $\beta(2)=\sqrt{8}$. We also study Brownian coagulation with a localized source and establish the behavior of the total cluster density and the total number of of clusters in the system. The latter quantity exhibits a logarithmic growth with time. 
\end{abstract}

\maketitle

\section{Introduction}

Aggregation underlies numerous phenomena ranging from polymerization \cite{Flory} and aerosol behavior \cite{skf} to formation of clouds \cite{cloud} and celestial matters where aggregation plays a significant role in planet and star formation (and perhaps even leads to compositeness of dark matter \cite{DM_14}). Earlier work on aggregation is described in \cite{SM17,C43}; the following work is summarized in a number of books and reviews \cite{agg-rev,OTB89,A99,FL03,book}. The framework for studying dilute well-mixed aggregating systems was developed by Smoluchowski~\cite{SM17} who wrote an infinite set of rate equations for the concentration $c_k(t)$ of clusters of ``mass'' $k$:
\begin{equation}
\label{Smol}
\frac{d c_k}{dt}=\frac{1}{2}\sum_{i+j=k}K_{i,j}c_ic_j-c_k\sum_{j\geq 1}K_{k,j}c_j
\end{equation}
These equations reflect that in applications each cluster is often composed of an integer number of identical elemental building blocks, `monomers'; then the number of constituting monomers can be called the mass of the cluster. Another tacit assumption is that the mass is the only relevant characteristic, so e.g. the shape of the cluster does not matter. This is certainly a simplification, although in some applications (e.g., when clusters are spherical or have some other fixed shape) it suffices to classify clusters only by their masses. 

There are only quadratic couplings between the densities in Eqs.~\eqref{Smol} since only binary collisions are taken into account. This is reasonable in dilute systems where binary interactions usually dominate \cite{biol:ternary}. The rate at which clusters of mass $i$ and $j$ merge is denoted by $K_{i,j}$. Mathematically, the reaction rates $K_{i,j}$ form an infinite symmetric matrix, $K_{i,j}=K_{j,i}$; the pre-factor $\tfrac{1}{2}$ in front of the gain terms on the right-hand side of \eqref{Smol} is needed to avoid double counting. 

In Brownian coagulation \cite{OTB89,book,SM17}, the reaction rate admits a simple expression, $K_{i,j}=4\pi(D_i+D_j)(R_i+R_j)$, through radii and diffusion coefficients of the merging droplets. In the regime when droplets are sufficiently small, the Stokes-Einstein relation implies $D_i\sim 1/R_i$, and the Brownian coagulation kernel becomes
\begin{equation}
\label{Brown:kernel}
K_{i,j}=\big(i^{1/3}+j^{1/3}\big)\big(i^{-1/3}+j^{-1/3}\big)
\end{equation}
Here we have dropped numerical factors and taken into account that $R_i\sim i^{1/3}$ for spherical droplet in three dimensions. Equations \eqref{Smol} with reaction kernel \eqref{Brown:kernel} are analytically intractable and Smoluchowski suggested to replace the Brownian coagulation kernel by the constant kernel, $K_{i,j}=\text{const}$. Equations \eqref{Smol} with this kernel are solvable. Apart from being the simplest possible kernel, the mass-independent kernel has the same homogeneity as the Brownian coagulation kernel \eqref{Brown:kernel}, namely in both cases $K_{\lambda i,\lambda j}=K_{i,j}$. Chief qualitative behaviors are similar in both cases, e.g., the typical cluster mass grows linearly with time: $\langle k\rangle \sim t$. (More subtle behaviors, e.g., the shape of the scaled mass distribution, differ  \cite{FL03,book}.) 

Aggregation with homogeneous source has also attracted considerable attention, see \cite{FL03,book,FS65,W82,HH87,T89,HT89,Colm12,JM98}. The details of the source play rather limited role, e.g., they do not affect the emerging large mass behavior if only clusters of small mass are injected. It is customary to assume that only monomers are injected. The governing equations then differ from \eqref{Smol} by the term $J\delta_{k,1}$, with $J$ being the strength of the monomer flux. Since the source is assumed to be homogeneous, the governing equations are still ordinary differential equations. Aggregating systems with homogeneous source often reach a non-equilibrium steady state \cite{FL03,book,FS65,W82,HH87,T89,HT89}; this makes such driven systems more tractable than un-driven systems. 

Here we study aggregation processes with a spatially localized source. The densities $c_k(\mathbf{r},t)$  vary according to a system of partial differential equations  (PDEs) 
\begin{eqnarray}
\label{loc:input}
\frac{\partial c_k}{\partial t} &=& D_k\nabla^2 c_k 
+ \frac{1}{2}\sum_{i+j=k}K_{i,j}c_ic_j-c_k\sum_{j\geq 1}K_{k,j}c_j \nonumber\\
&+&J \delta_{k,1}\delta(\mathbf{r})
\end{eqnarray}
The first term on the right-hand side describes mixing due to diffusion, the next two terms account for aggregation, and the last term represents the monomer source at the origin. 

At first sight, the system \eqref{loc:input} of infinitely many coupled non-linear PDEs is totally intractable. We are chiefly interested, however, in the large time behavior when cluster densities may become stationary. Whenever this happens, Eqs.~\eqref{loc:input} reduce to ordinary differential equations (ODEs). These ODEs are still generally intractable, so one can try to get insight from the simplest model with mass-independent reaction rates supplemented by mass-independent diffusion coefficients. 

The remainder of this paper is organized as follows. In Sec.~\ref{sec:const} we analyze the model in which reaction rates and diffusion coefficients are constant.  We begin with the total cluster density (asymptotic behaviors of the total cluster density far away from the source were originally studied in Ref.~\cite{Sid89}). We then derive more subtle decay laws for cluster densities. The results of Sec.~\ref{sec:const} show that the basic quantities decay algebraically with distance from the source, and suggest to seek similar solutions for models where reaction rates and diffusion coefficients are homogeneous functions of mass. In Sec.~\ref{sec:Brownian} we apply this strategy to Brownian coagulation with localized input. Finally, in Sec.~\ref{sec:conl}, we give a summary. 

\section{Mass-Independent Reaction Rates and Diffusion Coefficients}
\label{sec:const}

Here we consider the model with mass-independent reaction rates and diffusion coefficients. Starting with \cite{Sid89}, all subsequent work on aggregation processes with a localized source was done in the framework of this simplest model. (Studies of driven systems with three-particle aggregation \cite{PLK:3particle} and of aggregation-annihilation processes \cite{PLK:source} with a localized source also used this framework.) Furthermore, to avoid the challenge of treating infinitely many coupled non-linear PDEs,  the process was often investigated only on the level of coalescence \cite{Sid89,PLK12} when the distinction between individual clusters is  ignored, so there is just one reaction rate and just one diffusion coefficient. 

Even for aggregation processes  with mass-independent reaction rates and diffusion coefficients some basic quantities haven't been analyzed. For instance, the behavior of the monomer density was apparently unknown; we will show that it 
decays algebraically, $c_1 \sim r^{-(\sqrt{17}+1)/2}$, far away from the source (in three dimensions). To derive this and other results we start with the full time-dependent equations 
\begin{equation}
\label{source:simple}
\frac{\partial c_k}{\partial t} =\nabla^2 c_k 
+ \sum_{i+j=k}c_ic_j-2c_k c +J \delta_{k,1}\delta(\mathbf{r})
\end{equation}
Here $c(\mathbf{r},t)=\sum_{k\geq 1}c_k(\mathbf{r},t)$ is the total cluster density. Equations \eqref{source:simple}
follow from \eqref{loc:input} when $K_{i,j}=2$ and $D_k=1$. This choice makes formulas less cluttered, and it does not limit generality since it essentially amounts to choosing the units of length and time. Indeed, consider the general case of arbitrary mass-independent reaction rates and diffusion coefficients: $K_{i,j}=2K$ and $D_k=D$. If $L$ denotes the unit of length and $T$ denotes the unit of time, the dimensions of the parameters $D$ and $K$ are $[D]=L^2/T$ and $[K]=L^d/T$.  Thus we can use $(K/D)^{1/(d-2)}$ as a unit of length and $(K^2/D^d)^{1/(d-2)}$ as a unit of time \cite{2d_exception}. Measuring in these units is equivalent to setting $K=1$ and $D=1$. 


To appreciate the relative simplicity of Eqs.~\eqref{source:simple} we emphasize two its consequences. First, we notice that the mass density $M(\mathbf{r},t)=\sum_{k\geq 1}kc_k(\mathbf{r},t)$ satisfies a closed equation, the diffusion equation with a localized source:
\begin{equation}
\label{M:simple}
\frac{\partial M}{\partial t} =  \nabla^2 M + J \delta(\mathbf{r})
\end{equation}
Equation \eqref{M:simple} is already valid when the diffusion coefficients are mass-independent (the reaction terms do not affect the evolution as the aggregation process is manifestly mass-conserving). Second, the total cluster density also obeys a closed PDE
\begin{equation}
\label{ct:simple}
\frac{\partial c}{\partial t} =\nabla^2 c - c^2 + J\delta(\mathbf{r})
\end{equation}
The validity of Eq.~\eqref{ct:simple} relies on the mass-independence of both the reaction rates and diffusion coefficients. 

Equation \eqref{ct:simple} is a non-linear PDE which has not been solved, but the asymptotic behaviors which are valid far away from the source are known \cite{Sid89}. We now outline these asymptotic behaviors and briefly discuss the matching with near-source behaviors. These results will be useful once we turn to the analysis of the full system \eqref{source:simple}. 

Let us start with the most relevant three-dimensional case. In the long time limit the total density should become time-independent. In the stationary regime in three dimensions we need to solve an ODE
\begin{equation}
\label{cr3}
\frac{1}{r^2}\,\frac{d}{dr}\left(r^2\,\frac{dc}{dr}\right) - c^2 + J\delta(\mathbf{r})=0
\end{equation}
The source term vanishes for all $r>0$. Away from the origin, Eq.~\eqref{cr3} reads $c''+\frac{2}{r}c'=c^2$ and admits a simple exact solution
\begin{equation}
\label{cr3:simple}
c = \frac{2}{r^2}
\end{equation}
This is just a special solution, not the general solution of the second order differential equation \eqref{cr3}. The source term actually prescribes the boundary condition, 
\begin{equation}
\label{BC:3d}
\lim_{r\to 0} 4\pi r^2 \frac{dc}{dr} = -J,
\end{equation}
and \eqref{cr3:simple} disagrees with this boundary condition. Equation \eqref{cr3:simple} yields the leading asymptotic far away from the source. Near the source, one anticipates that the reaction term is negligible and hence $c=J/(4\pi r)$. One straightforwardly finds further corrections:
\begin{equation}
\label{cr3:near}
c = \frac{J}{4\pi r}+\left(\frac{J}{4\pi}\right)^2 \ln (Jr)+ \ldots
\end{equation}

Equating the two asymptotic behaviors, $r^{-2}\sim J r^{-1}$, we estimate the location of the crossover region: $r\sim J^{-1}$. The asymptotic \eqref{cr3:simple} [resp. \eqref{cr3:near}] is therefore applicable when $r\gg J^{-1}$ [resp. $r\ll J^{-1}$].

The location of the crossover can be alternatively established without knowing the asymptotic behaviors. Indeed, the governing equation \eqref{cr3} and the boundary condition \eqref{BC:3d} are invariant under the one-parameter transformation group 
\begin{equation}
r\to \lambda r, \quad c\to \lambda^{-2}c, \quad J\to \lambda^{-1}J
\end{equation}
Therefore the solution to Eq.~\eqref{cr3} must have the form
\begin{equation}
\label{inv:3}
c(r) = J^2  \mathcal{C}(R), \quad R=Jr
\end{equation}
The scaled density $\mathcal{C}(R)$ is universal (that is, independent on $J$) and it satisfies
\begin{equation}
\label{Cr3}
\frac{1}{R^2}\,\frac{d}{dR}\left(R^2\,\frac{d\mathcal{C}}{dR}\right) - \mathcal{C}^2 + \delta(\mathbf{R})=0
\end{equation}
The crossover between near and far-from-source region must occur when $R\sim 1$, i.e., when $r\sim J^{-1}$.

In terms of the scaled distance, the asymptotic behaviors of the scaled density are
\begin{equation}
\label{cr3:scaled}
\mathcal{C} \simeq
\begin{cases}
1/(4\pi R)     & R\ll 1\\
2/R^2           & R\gg 1
\end{cases}
\end{equation}

The universality of the asymptotic behavior \eqref{cr3:simple} is remarkable---it does not depend on the strength of the source. Thus the details of input play no role at all as far as the leading asymptotic behavior of the total density is concerned; near the source the details become relevant. In the most general case when various cluster species are injected, so the source is characterized by $\{J_k| ~k=1,2,\ldots\}$, the far-away asymptotic is still given by  the same formula \eqref{cr3:simple}, while near the source we can use \eqref{cr3:near} with $J=\sum_{k\geq 1}J_k$.

Consider now our driven aggregation process \eqref{source:simple} in arbitrary spatial dimension $d$. The analog of \eqref{cr3:simple} reads 
\begin{equation}
\label{crd:simple}
c = \frac{2(4-d)}{r^2}
\end{equation}
The analytical solution is again unavailable. The invariance of the governing equation allows us to write
\begin{equation}
\label{inv:d}
c(r) = J^{2/(4-d)}  \mathcal{C}(R), \quad R=J^{1/(4-d)} r
\end{equation}
The asymptotic behaviors of the scaled density are
\begin{equation}
\mathcal{C} \simeq
\label{crd:scaled}
\begin{cases}
1/[(d-2)\Omega_d\, R^{d-2}]  & R\ll 1\\
2(4-d)/R^2                              & R\gg 1
\end{cases}
\end{equation}
where $\Omega_d=2\pi^{d/2}/\Gamma(d/2)$ is the surface ``area'' of the unit sphere $\mathbb{S}^{d-1}$ in $d$ dimensions.

These results are valid only when $d<4$. The vanishing of the amplitude in \eqref{crd:simple} at $d=4$ suggests a possible logarithmic correction, and it is indeed easy to establish \cite{book} the asymptotic behavior far away from the source
\begin{equation}
\label{cr4:simple}
c\simeq \frac{2}{r^2\ln r}
\end{equation}
Note that the cluster density is still independent on the strength $J$ of the source. The crossover from the far away asymptotic \eqref{cr4:simple} to the near-source behavior $c\simeq \frac{J}{4\pi^2 r^2}$ occurs when $r\propto \exp(8\pi^2/J)$. 

For $d>4$, the reaction term is negligible far away from the source \cite{Sid89}. Close to the source, the density is found by solving $\nabla^2 c + J\delta(\mathbf{r})=0$ leading to 
\begin{equation}
\label{cr5:simple}
c\simeq \frac{J}{(d-2)\Omega_d\,r^{d-2}}\qquad {\rm when}\quad d>4
\end{equation}
Far away from the source, the governing equation is asymptotically the same, but the source should be renormalized. The solution thus has the same form \eqref{cr5:simple}, with a renormalized source strength $J_\text{eff}$.  The reaction terms are important when $r\sim J^{1/(d-4)}$.

An extra care is required in dimensions $d\leq 2$. The rate equations \eqref{loc:input} are valid above the critical dimension. For the constant-kernel aggregation, the critical dimension is $d_c=2$ (see e.g. \cite{book}). In two dimensions,  the correct behavior is obtained after a mild modification of the rate equations, e.g., \eqref{ct:simple} should be replaced by 
\begin{equation}
\label{ct2:simple}
\frac{\partial c}{\partial t} =\nabla^2 c - c^2/\ln(1/c) + J\delta(\mathbf{r})
\end{equation}
from which 
\begin{equation}
\label{cr2:simple}
c\simeq \frac{8 \ln r}{r^2}
\end{equation}
far away from the source. 
 
Collecting the asymptotic behaviors far away from the source and adding the one-dimensional asymptotic (see \cite{Sid89}), we get
\begin{equation}
\label{cr:simple}
c\sim
\begin{cases}
r^{-1}                   & d=1\\
r^{-2} \ln r            & d=2\\
r^{-2}                   & d=3\\
r^{-2} (\ln r)^{-1}  & d=4\\
J r^{-(d-2)}           & d>4
\end{cases}
\end{equation}

The total number of clusters $\mathcal{N}(t)=\int d\mathbf{r}\,c(\mathbf{r},t)$ also exhibits amusing behaviors. The density quickly vanishes beyond the diffusive length $r\sim\sqrt{t}$. Thus we use \eqref{cr:simple} and perform the integration up to $r\sim\sqrt{t}$ to find \cite{J:eff}
\begin{equation}
\label{Nt:simple}
\mathcal{N} \sim
\begin{cases}
\ln t              & d=1\\
(\ln t)^2        & d=2\\
\sqrt{t}         & d=3\\
t/\ln t            & d=4\\
J t                & d>4
\end{cases}
\end{equation}

Spatial behaviors \eqref{cr:simple} for the total cluster density were established in Ref.~\cite{Sid89}. In Eqs.~\eqref{cr:simple}--\eqref{Nt:simple} we additionally emphasized a rather paradoxical lack of dependence on $J$ in $d\leq 4$ dimensions.  Thus the entire process is driven by the source, yet the leading behavior of the total number of clusters in the system is insensitive to the strength $J$ as long as the source is non-vanishing. 

Little is known on individual cluster densities. We now establish some of their properties. 

\subsection{Monomer Density}

The asymptotic behaviors of the monomer density interestingly vary with spatial dimension: 
\begin{equation}
\label{mon:simple}
c_1\sim
\begin{cases}
r^{-5}                            & d=1\\
J^{-(\sqrt{2}-1)}\,r^{-\sqrt{8}}                   & d=2\\
J^{-(\sqrt{17}-3)/2}\,r^{-(\sqrt{17}+1)/2}   & d=3\\
J^{-1}\,r^{-2} (\ln r)^{-2}                           & d=4\\
J\,r^{-(d-2)}                                              & d>4
\end{cases}
\end{equation}
In one dimension, the analysis \cite{Sid89,PLK:source} requires separate techniques (outlined below in Sec~\ref{sec:d1}) as the rate equation framework is inapplicable. The higher-dimensional behavior is simple to appreciate. Indeed, above four dimensions clusters do not ``see'' each other far away from the source and hence $c_k\sim r^{-(d-2)}$ is valid for all $k$. The four-dimensional case is most tractable and we shall determine the entire cluster mass distribution (Sec.~\ref{d4}). In two and three dimensions, the exponent describing the large distance behavior is irrational. 

We now deduce these unusual behaviors from the governing equations. When $r\ll \sqrt{t}$, the monomer density is stationary. Thus in three dimensions we must solve
\begin{equation}
\label{mon3:simple}
\nabla^2 c_1 -  2c_1 c + J\delta(\mathbf{r})=0
\end{equation}
Recalling that $c=2/r^2$ in three dimensions far away from the source, we recast \eqref{mon3:simple} into
\begin{equation}
\label{m3:simple}
\frac{1}{r^2}\,\frac{d}{dr}\left(r^2\,\frac{dc_1}{dr}\right)=\frac{4c_1}{r^2}
\end{equation}
The general solution of this linear ODE is a combination of pure algebraic functions, $r^{\beta_\pm}$, with $\beta_\pm=-\frac{1}{2}\pm \frac{1}{2}\sqrt{17}$.  To ensure that the density does not diverge as $r\to\infty$, we pick $c_1 = A\, r^{-(\sqrt{17}+1)/2}$. Thus we arrive at the announced irrational exponent. It is impossible to deduce the dependence of the amplitude $A$ on the strength $J$ of the source from a {\em linear} Eq.~\eqref{m3:simple}. However, we can determine the amplitude on the general grounds, viz. using the invariance of the governing equation  \eqref{mon3:simple}. Equations \eqref{cr3} and \eqref{mon3:simple} are invariant under the transformation group 
\begin{equation}
r\to \lambda r, \quad c\to \lambda^{-2}c,  \quad c_1\to \lambda^{-2}c_1, \quad J\to \lambda^{-1}J
\end{equation}
and therefore
\begin{equation}
\label{inv:31}
c(r) = J^2  \mathcal{C}(R), \quad c_1(r) = J^2  \mathcal{C}_1(R), \quad R=Jr
\end{equation}
{}From $c_1 \propto r^{-(\sqrt{17}+1)/2}$ we get $\mathcal{C}_1 \sim R^{-(\sqrt{17}+1)/2}$ leading to $c_1 \propto J^{-(\sqrt{17}-3)/2}$.

In two dimensions, we must solve
\begin{equation}
\label{mon2:simple}
\nabla^2 c_1 =  \frac{2c_1 c}{\ln(1/c)} 
\end{equation}
Using \eqref{cr2:simple} and keeping only the leading term we get
\begin{equation}
\label{m2:simple}
\frac{1}{r}\,\frac{d}{dr}\left(r\,\frac{dc_1}{dr}\right)=\frac{8c_1}{r^2}
\end{equation}
leading to the announced asymptotic $c_1 \sim r^{-\sqrt{8}}$. The dependence on $J$ appearing in $c_1\sim J^{-(\sqrt{2}-1)}\,r^{-\sqrt{8}}$ is established using the invariant form of the solution:
\begin{equation}
\label{inv:21}
c(r) = J \mathcal{C}(R), \quad c_1(r) = J \mathcal{C}_1(R), \quad R=r\sqrt{J}
\end{equation}

Treating the dimensionality $d$ as a continuous parameter we find that in the general case $2\leq d<4$ the density of monomers far away from the source obeys
\begin{equation}
c_1\sim J^{-\alpha(d)}\,r^{-\beta(d)}
\end{equation}
The exponent $\beta(d)$ is found by solving $\nabla^2 c_1 = 2c_1 c$, or
\begin{equation*}
\frac{1}{r^{d-1}}\,\frac{d}{dr}\left(r^{d-1}\,\frac{dc_1}{dr}\right)=\frac{4(4-d)c_1}{r^2}
\end{equation*}
to yield
\begin{equation}
\beta(d)   = \frac{d-2+\sqrt{(d-2)^2+16(4-d)}}{2}
\end{equation}
Another exponent 
\begin{equation}
\alpha(d) = \frac{\beta(d)-2}{4-d}
\end{equation}
is determined using the invariant form of the solution,
\begin{equation*}
c_1(r) = J^{2/(4-d)}  \mathcal{C}_1(R), \quad R=J^{1/(4-d)} r,
\end{equation*}
together with $\mathcal{C}_1 \sim R^{-\beta(d)}$.

\subsection{Mass Distribution in $d=4$}
\label{d4}

The marginal case of $d=4$ is the simplest and all densities $c_k(r)$ can be determined analytically. Let us start with monomers. In the stationary regime, they satisfy $\nabla^2 c_1 =  2c_1 c$, so in four dimensions 
\begin{equation}
\label{m4:simple}
\frac{1}{r^3}\,\frac{d}{dr}\left(r^3\,\frac{dc_1}{dr}\right)=\frac{4c_1}{r^2 \ell}
\end{equation}
where we used \eqref{cr2:simple} and the shorthand notation $\ell = \ln r$. Solving \eqref{m4:simple} gives the leading behavior 
\begin{equation}
\label{m4:sol}
c_1 = \frac{A}{r^{2}\,\ell^{2}}
\end{equation}
with yet undetermined amplitude $A$. Similarly for dimers the governing equation reads
\begin{equation}
\label{d4:simple}
\frac{1}{r^3}\,\frac{d}{dr}\left(r^3\,\frac{dc_2}{dr}\right)=\frac{4c_2}{r^2 \ell} - c_1^2
\end{equation}
Plugging \eqref{m4:sol} into \eqref{d4:simple} and solving for the density of dimers we find
\begin{equation}
\label{d4:sol}
c_2 = \frac{A}{r^{2}\,\ell^{2}}
\end{equation}
with the same amplitude as in \eqref{m4:sol}. Solving for a few more densities we guess the structure of the solution:
\begin{equation}
\label{ck4:sol}
c_k = \frac{2}{r^2}\,C_k(\ell)
\end{equation}
Plugging this ansatz into
\begin{equation}
\frac{1}{r^3}\,\frac{d}{dr}\left(r^3\,\frac{dc_k}{dr}\right) + \sum_{i+j=k}c_ic_j = 2c_k c
\end{equation}
we arrive at \cite{asymp} 
\begin{equation}
\label{CK}
\frac{d C_k}{d\ell}=\sum_{i+j=k}C_i C_j- 2C_k C
\end{equation}
where $C=\sum_{k\geq 1}C_k$ is asymptotically equal to $\ell^{-1}$. Equation \eqref{CK} is {\em identical} to the solvable Smoluchowski for the constant-kernel aggregation---the variable $\ell = \ln r$ plays the role of time. The solution reads 
\begin{equation}
\label{CK:scaling}
C_k = \frac{A}{\ell^2}\,\exp\!\left(-\frac{Ak}{\ell}\right)
\end{equation}
and it manifestly satisfies the sum rule $\sum_{k\geq 1}C_k=\ell^{-1}$. To fix $A$ we recall that the mass density $M=\sum_{k\geq 1}kc_k$ satisfies \eqref{M:simple} which admits an asymptotically stationary solution when $d>2$. For $d=4$, we have
\begin{equation}
\label{M4:simple}
M = \frac{J}{4\pi^2 r^2}\,,
\end{equation}
which in conjunction with \eqref{ck4:sol} yields the sum rule
\begin{equation*}
\sum_{k\geq 1} kC_k=  \frac{J}{8\pi^2}
\end{equation*}
fixing the amplitude $A=8\pi^2/J$. 

Combining previous findings we write the mass distribution in four dimensions 
\begin{equation}
\label{ck4:final}
c_k = \frac{16\pi^2}{J}\,\frac{1}{r^2\,\ell^2}\,\exp\!\left(\!-\frac{8\pi^2}{J}\,\frac{k}{\ell}\right)
\end{equation}

\subsection{Mass Distribution in $d=3$}

Let us immediately seek the mass distribution in the scaling form
\begin{equation}
\label{ck3:sol}
c_k(r) = r^{-3}\,F(x), \quad x = \frac{k}{r}
\end{equation}
In the scaling regime, clusters are typically large, so we must require $r\gg J^{-1}$ to assure that the reaction term dominates over the source term. 

The choice of the scaling form \eqref{ck3:sol} is suggested by the sum rules 
\begin{equation}
\label{s_rules}
\sum_{k\geq 1}c_k(r)=\frac{2}{r^2}\,, \qquad \sum_{k\geq 1}kc_k(r)= \frac{J}{4\pi\,r}
\end{equation}
The second sum rules merely gives the mass density which is the solution of Eq.~\eqref{M:simple} in $d=3$. The sum rules \eqref{s_rules} are manifestly obeyed and they imply two integral relations on the scaling function $F(x)$:
\begin{equation}
\label{s_rules:F}
\int_0^\infty dx\,F(x) = 2, \quad \int_0^\infty dx\,xF(x) = \frac{J}{4\pi}
\end{equation}

Plugging \eqref{ck3:sol} into the governing equations yields
\begin{equation}
\label{F:eq}
x^2 \frac{d^2 F}{dx^2} + 6x \frac{d F}{dx} + 2F + \int_0^x dy\,F(y)F(x-y) = 0
\end{equation}
The integral is the convolution. Therefore we make the Laplace transform, $\Phi(s) = \int_0^\infty dx\,e^{-xs} F(x)$, and recast \eqref{F:eq} into an ordinary differential equation 
\begin{equation}
\label{Phi:eq}
s^2 \Phi'' - 2s\Phi' - 2\Phi + \Phi^2 = 0
\end{equation}
Here the prime denotes the differentiation with respect to $s$. The sum rules \eqref{s_rules:F} become
\begin{equation}
\label{s_rules:Phi}
\Phi(0) = 2, \quad \Phi'(0) = - \frac{J}{4\pi}
\end{equation}

It appears impossible to find an analytic solution of \eqref{Phi:eq} subject to \eqref{s_rules:Phi}. An asymptotic analysis gives
\begin{subequations}
\begin{align}
&F\sim x^{(\sqrt{17}-5)/2} \qquad x\to 0
\label{F:small} \\
&F\simeq \frac{J^2}{16\pi^2}\,\, x^{-3} \qquad ~x\to \infty
\label{F:large} 
\end{align}
\end{subequations}
The $x\to 0$ behavior of $F(x)$ can be extracted from the $s\to\infty$ behavior of $\Phi(s)$. Since $\Phi(\infty)=0$, we can drop the non-linear term in \eqref{Phi:eq} when $s\to\infty$. Solving the linear equation we get $\Phi\sim s^{-(\sqrt{17}-3)/2}$, from which we deduce \eqref{F:small}. The $x\to \infty$ behavior of $F(x)$ is incapsulated in the $s\to 0$ behavior of $\Phi(s)$. The latter is deduced from \eqref{Phi:eq} and \eqref{s_rules:Phi} by straightforward manipulations: 
\begin{equation*}
\Phi = 2 - \frac{J}{4\pi}\,s -  \frac{J^2}{32\pi^2}\,s^2\ln s + \ldots
\end{equation*}
This small $s$ expansion yields \eqref{F:large}.

Combining the asymptotic behaviors \eqref{F:small}--\eqref{F:large} with \eqref{ck3:sol} we arrive at the asymptotic behaviors of the mass distribution in three dimensions:
\begin{subequations}
\begin{align}
&c_k(r)\sim k^{(\sqrt{17}-5)/2}\, r^{-(\sqrt{17}+1)/2} ~\quad k\ll r
\label{small} \\
&c_k(r)\simeq \frac{J^2}{16\pi^2}\, k^{-3}  \quad \qquad \qquad \qquad  k\gg r
\label{large} 
\end{align}
\end{subequations}
We emphasize that these results are valid only far away from the source: $r\gg J^{-1}$. 

\subsection{Mass Distribution in $d=1$}
\label{sec:d1}

A lot is known in one dimension \cite{Sid89,PLK:source,FL03,book}. Here we recall a few basic results and provide more details for the behavior at the origin which is rather peculiar: The densities turn out to be are finite (in higher dimensions, they are infinite). 

In the stationary regime, the governing equation for the total cluster density 
\begin{equation}
\label{cr1}
\frac{d^2 c}{dx^2} - c^2 + J\delta(x)=0
\end{equation}
is solvable \cite{Sid89}
\begin{equation}
\label{c:1d}
c(x) = \frac{6}{(|x|+\ell)^2}\,, \quad \ell = \left(\frac{24}{J}\right)^{1/3}
\end{equation}
The structure of \eqref{c:1d} agrees with the general form \eqref{inv:d} and it gives the explicit form, $\mathcal{C}(X)=6/\big[|X|+(24)^{1/3}\big]^2$ with $X=J^{1/3}x$, of the scaled density in one dimension. 

The stationary cluster densities are encapsulated in the generating function \cite{Sid89,PLK:source,FL03}
\begin{equation}
\label{GF:1d}
\sum_{k\geq 1} (1-z^k)c_k(x) = \frac{6}{(|x|+\ell(1-z)^{-1/3})^2}
\end{equation}

The densities remain finite at the origin. Specializing \eqref{GF:1d} to $x=0$ and expanding in powers of $z$ one finds
\begin{equation}
\label{ck:origin}
c_k(0) = \left(\frac{J}{3}\right)^{2/3} \frac{\Gamma\big(k-\frac{2}{3}\big)}{\Gamma\big(\frac{1}{3}\big) \Gamma(k+1)}
\end{equation}

The above results, e.g. the exact total density \eqref{c:1d}, or following from  \eqref{GF:1d} exact cluster densities 
\begin{equation*}
\begin{split}
c_1(x) & = \frac{4\ell}{(|x|+\ell)^3}\\
c_2(x) & = \frac{8}{3}\, \frac{\ell}{(|x|+\ell)^3} - \frac{2\ell^2}{(|x|+\ell)^4}\\
c_3(x) & =  \frac{56}{27}\, \frac{\ell}{(|x|+\ell)^3} - \frac{8}{3}\, \frac{\ell^2}{(|x|+\ell)^4} 
+ \frac{8}{9}\,\frac{\ell^3}{(|x|+\ell)^5}
\end{split}
\end{equation*}
etc. disagree with asymptotic predictions, e.g., with far from the source asymptotic behaviors $c\sim |x|^{-1}$ and $c_1\sim |x|^{-5}$ appearing in \eqref{cr:simple} and \eqref{mon:simple}. This is because the above exact results have been established in the rate equation framework, while in Eqs.~\eqref{cr:simple} and \eqref{mon:simple} we cited predictions for the truly one-dimensional process. For our aggregation process, the rate equations are valid only for $d>2$. This does not imply, however, that the above exact results are useless. Indeed, if the aggregation process occurs in three dimensions and new particles are injected on the two-dimensional interface, the rate equation  framework is valid. 

We now illustrate the techniques which are used for the treatment of the truly one-dimensional process. These  techniques are rather special, see \cite{T89,Sid89,PLK:source,book}, and they are ultimately based on the assumption that the diffusion coefficients are mass-independent. Another crucial postulate is the assumption that clusters are size-less (point particles if the process occurs on the line, or each cluster occupies a single lattice point). We also note that in one dimension, it is rather natural to examine particles undergoing biased motion, and in this situation one often needs even more complicated methods, see \cite{Asymmetric:source,Kirone:source}. We always consider symmetric motion.

A powerful treatment of the one-dimensional aggregation process with mass-independent diffusion coefficients is based \cite{T89} on functions $P_k(x,y;t)$, where $x<y$ and $k=0,1,2,\ldots$. By definition, $P_k(x,y;t)$ is the probability that the total mass contained in the interval $[x,y]$ at time $t$ is equal to $k$. Only the total mass is taken into account: For $k=3$, for instance, there could be three monomers, or one monomer and one dimer, but these details (and the precise locations of the clusters in the interval $[x,y]$) do not enter into the description. The probabilities $P_k(x,y;t)$ evolve according to 
\begin{eqnarray}
\label{Pkxyt}
&&\frac{\partial }{\partial t}\,P_k(x,y; t)= 
\left(\frac{\partial^2 }{\partial x^2}+\frac{\partial^2 }{\partial y^2}\right)P_k(x,y; t)\nonumber\\
&& + J [P_{k-1}(x,y; t) - P_k(x,y; t)]\theta(y)\theta(-x)
\end{eqnarray}
with Heaviside step functions assuring that the source contributes only when $x<0<y$. Equations \eqref{Pkxyt} are linear and solvable, although the solution is rather cumbersome. In contrast to higher dimensions where the governing PDEs are non-linear and non-solvable, in one dimension one can obtain exact results for the full time-dependent behavior. 

Let us look at the stationary regime and additionally focus on a symmetric (in space) subset of probabilities: $Q_k(x)=P_k(-x,x;t=\infty)$. Equations \eqref{Pkxyt} give
\begin{equation}
\label{Qkx}
0 = 2\,\frac{d^2 Q_k}{d x^2} + J[Q_{k-1}-Q_k]
\end{equation}
The generating function $Q(x,z)=\sum_{k\geq 0} z^k Q_k(x)$ satisfies
\begin{equation}
\label{Qxz}
2\,\frac{d^2 Q}{d x^2} = J(1-z)Q
\end{equation}
Solving this equation subject to the boundary conditions
\begin{equation}
\label{Q:BC}
Q(x=\infty,z)=0, \quad Q(x=0,z)=1
\end{equation}
we obtain
\begin{equation}
\label{Q:sol}
Q(x,z)=\exp\!\left[ -x \sqrt{\frac{J}{2}}\, \sqrt{1-z} \right]
\end{equation}
The densities at the origin are $c_k(0)=\frac{1}{2}\frac{d Q_k}{d x}\big|_{x=0}$. Hence
\begin{equation}
\sum_{k\geq 0} z^k c_k(0) = \frac{1}{2}\,\frac{d Q}{d x}\Big|_{x=0}= - \sqrt{\frac{J}{8}}\, \sqrt{1-z}
\end{equation}
from which
\begin{equation}
c_k(0) = \sqrt{\frac{J}{32\pi}}\,\,\frac{\Gamma\big(k-\frac{1}{2}\big)}{\Gamma(k+1)}
\end{equation}
Pushing these calculations one can determine the spatial distributions $c_k(x)$.

\section{Brownian Coagulation with Spatially Localized Source}
\label{sec:Brownian}

In this section we study Brownian coagulation driven by a localized source of monomers. We begin with the classical Brownian coagulation when clusters are spherical \cite{SM17,C43,agg-rev,OTB89,book,van,vD89}. In three dimensions the reaction rates are given by \eqref{Brown:kernel} and the diffusion coefficients are
\begin{equation}
\label{Brown:diff}
D_k =  k^{-1/3}
\end{equation}
Generally $K_{i,j}\sim (D_i+D_j)(R_i+R_j)^{d-2}, ~D_k\sim R_k^{-(d-2)}$ in $d>2$ dimensions, so (ignoring numerical factors) the reaction rates are 
\begin{equation}
\label{Brown:kernel_d}
K_{i,j}=\big(i^{1/d}+j^{1/d}\big)^{d-2}\big(i^{-1+2/d}+j^{-1+2/d}\big)
\end{equation}
and the diffusion coefficients are
\begin{equation}
\label{Brown:diff_d}
D_k =  k^{-1+2/d}
\end{equation}
for spherical clusters undergoing Brownian coagulation in $d>2$ dimensions. 

We then briefly explore what can happen when clusters are fractal. In three dimensions we still can use the general prediction $K_{i,j}\sim (D_i+D_j)(R_i+R_j)$ of the reaction rate theory \cite{SM17,OTB89,book} and the Stokes-Einstein relation $D_k\sim 1/R_k$. The characteristic size of a fractal cluster grows with its mass as $R_k\sim k^a$, where $a=1/D_f$ is the inverse fractal dimension of clusters. Thus in three dimensions, the generalized Brownian kernel is 
\begin{equation}
\label{Brown:3d-K}
K_{i,j}= (i^a+j^a)(i^{-a}+j^{-a})
\end{equation}
and the diffusion coefficients are
\begin{equation}
\label{Brown:3d-D}
D_k =  k^{-a}
\end{equation}

\subsection{Classical Brownian Coagulation}

Consider the classical Brownian coagulation and assume that the densities becomes stationary in the long time limit. The  governing equations read
\begin{equation}
\label{Brown:ck}
D_k\nabla^2 c_k 
+ \frac{1}{2}\sum_{i+j=k}K_{i,j}c_ic_j-c_k\sum_{j\geq 1}K_{k,j}c_j = 0
\end{equation}
for $k\geq 2$. The density of monomers satisfies
\begin{equation}
\label{Brown:c1}
D_1\nabla^2 c_1 - c_1\sum_{j\geq 1}K_{1,j}c_j + J\delta(\mathbf{r})= 0
\end{equation}

This infinite system of coupled non-linear ODEs with reaction kernel \eqref{Brown:kernel_d}  and diffusion coefficients \eqref{Brown:diff_d} is analytically intractable. Let us try to establish major features using heuristic arguments.

Summing all Eqs.~\eqref{Brown:ck} and \eqref{Brown:c1} we obtain
\begin{equation}
\label{c:steady}
\nabla^2 \sum_{k\geq 1} D_kc_k 
- \frac{1}{2}\sum_{i, j \geq 1}K_{i,j}c_ic_j + J\delta(\mathbf{r})= 0
\end{equation}
Let $s=s(r)$ be a typical mass. The typical diffusion coefficient is $D_s=s^{-1+2/d}$. Invoking scaling we estimate $\sum_{k\geq 1} D_kc_k\sim s^{-1+2/d} c$. The kernel \eqref{Brown:kernel_d} has homogeneity index zero since it satisfies $K_{\lambda i,\lambda j}=K_{i,j}$. Therefore we replace the Brownian \eqref{Brown:kernel_d} by constant kernel in scaling estimates. This gives $\sum_{i, j \geq 1}K_{i,j}c_ic_j \sim c^2$, so Eq.~\eqref{c:steady} leads to relation
\begin{equation}
\label{cs1:steady}
r^{-2} s^{-1+2/d} c \sim c^2
\end{equation}

To establish another relation we multiply \eqref{Brown:ck} by $k$, sum over all $k\geq 2$, and also add \eqref{Brown:c1} to give
\begin{equation}
\label{M:steady}
\nabla^2 \sum_{k\geq 1} kD_kc_k + J\delta(\mathbf{r})= 0
\end{equation}
from which 
\begin{equation}
\label{M_sol:steady}
\sum_{k\geq 1} kD_kc_k =\frac{J}{(d-2)\Omega_d}\,\frac{1}{r^{d-2}}
\end{equation}
where $\Omega_d=2\pi^{d/2}/\Gamma(d/2)$. On the level of scaling we estimate $\sum_{k\geq 1} kD_kc_k\sim s^{2/d} c$. Therefore Eq.~\eqref{M_sol:steady} implies
\begin{equation}
\label{cs2:steady}
s^{2/d} c \sim \frac{J}{r^{d-2}}
\end{equation}
From relations \eqref{cs1:steady} and \eqref{cs2:steady} we deduce
\begin{equation}
\label{cs:steady}
c \sim r^{-d} J^{-(d-2)/(4-d)}\,, \quad s \sim r^{d} J^{d/(4-d)}
\end{equation}
Varying $d$ and looking at the dependence on $J$ we realize that the above results are valid when $2<d<4$. The lower bound is already evident from Eq.~\eqref{M_sol:steady}, the stationary solution \eqref{M_sol:steady} exists only when $d>2$. The upper bound emerges from \eqref{cs:steady}. For $d>4$ there is no scaling, e.g. a finite fraction of monomers escape to infinity without participating in aggregation events. 

The total number of clusters $\mathcal{N}(t)=\int d\mathbf{r}\,c(\mathbf{r},t)$ is estimated by using \eqref{cs:steady} to give 
\begin{equation*}
\mathcal{N}(t)\sim \int_0^R dr\, r^{d-1}c(r)\sim J^{-(d-2)/(4-d)} \int_0^R \frac{dr}{r}
\end{equation*}

The integral varies logarithmically with time. One can make a more precise estimate of the cutoff length $R$. We write $R\sim \sqrt{Dt}$ with $D\sim s^{-1+2/d}$ [see Eq.~\eqref{Brown:diff_d}] leading to $R^2\sim s^{-1+2/d} t$. Further, Eq.~\eqref{cs:steady} yields $s \sim R^{d} J^{d/(4-d)}$. From these two relations we obtain
\begin{equation}
R\sim \left[J^{-(d-2)/(4-d)}\,t\right]^{1/d}
\end{equation}
Thus
\begin{equation}
\mathcal{N}(t)\sim J^{-(d-2)/(4-d)} \,\ln\!\big[J^{-(d-2)/(4-d)}\,t\big]
\end{equation}

In three dimensions our main scaling results read
\begin{equation}
\label{csN}
c\sim J^{-1}\, r^{-3}\,, \quad 
s\sim J^{3}\, r^{3}\,, \quad 
\mathcal{N} \sim J^{-1}\, \ln(t/J)
\end{equation}
These scaling laws imply that the mass distribution in three dimensions has the scaling form
\begin{equation}
\label{ckr}
c_k(r) = J^{-4}\, r^{-6} F(x), \quad x = \frac{k}{J^{3}\, r^{3}}
\end{equation}

It appears impossible to extract quantitative information even about limiting behaviors of the scaled mass distribution $F(x)$. For instance, one would like to establish how the monomer density decays with distance. We estimate
\begin{equation*}
\sum_{j \geq 1}K_{1,j}c_j \sim \sum_{j \geq 1} j^{1/3} c_j \sim s^{1/3} c\sim r^{-2}
\end{equation*}
and hence Eq.~\eqref{Brown:c1} gives $\nabla^2 c_1 \sim r^{-2} c_1$. This equation is invariant under the scale transformation, $r\to \lambda r$, and hence the solution is algebraically decaying: $c_1\sim r^{-\beta}$. To determine the decay exponent, however, we need to know the precise numerical factor which cannot be obtained in the framework of the heuristic approach used in this section. 

\subsection{Generalized Brownian Coagulation}

If clusters are fractal, the proper reaction rates and diffusion coefficients are given by \eqref{Brown:3d-K}--\eqref{Brown:3d-D}. Various fractal dimensions, and respective values of the parameter $a=1/D_f$ in Eqs.~\eqref{Brown:3d-K}--\eqref{Brown:3d-D}, appear in applications. In polymerization \cite{PP} if emerging polymers are linear chains, the value $a=1$ describes the situation when the polymers are stiff (so their fractal dimension is $D_f=1$). Ideal polymer chains (also known as polymers in the $\theta$ solvent) are essentially random walks, so $D_f=2$ and $a=\frac{1}{2}$. Polymers in a good solvent are essentially self-avoiding walks, so $a\approx \frac{3}{5}$. 

Assuming that the stationary regime is reached, we perform the same analysis as in the previous subsection and obtain Eq.~\eqref{c:steady} leading to $r^{-2} s^{-a} c \sim c^2$. Equations \eqref{M:steady}--\eqref{M_sol:steady} still hold and from the latter we deduce another relation: $s^{1-a} c \sim J/r$ in three dimensions. Thus the cluster density and the typical mass are
\begin{equation}
\label{cs:a}
c\sim J^{-a/(1-2a)}\, r^{-(2-3a)/(1-2a)}\,, \quad 
s\sim (Jr)^{1/(1-2a)}
\end{equation}
Using the former result we find that the total number of clusters behaves according to
\begin{equation}
\label{Nt:GB}
\mathcal{N}(t)\sim J^{-1} \times
\begin{cases}
(J^2 t)^{(1-3a)/(2-3a)}  & a <\frac{1}{3}\\
\ln (t/J)                         &  a=\frac{1}{3}\\
\text{const}                  &  \frac{1}{3}< a 
\end{cases}
\end{equation}

In three dimensions $D_f\leq 3$ and hence $a\geq \frac{1}{3}$, so in the physically relevant cases we anticipate that the total number of clusters either grows logarithmically with time ($a=\frac{1}{3}$), or saturates ($\frac{1}{3}< a$). 

Equation \eqref{cs:a} shows that the exponents diverge when $a\to \frac{1}{2}$ from below. Further, the predictions Eq.~\eqref{cs:a} for $a>\frac{1}{2}$ are dubious. These are the indications that the stationary regime is never reached when $a\geq \frac{1}{2}$. A similar phenomenon, namely that three-dimensional Brownian coagulation with reaction kernel \eqref{Brown:3d-K} never reaches a steady state when $a>\frac{1}{2}$, has been established in the case when the source was homogeneous \cite{Colm12}. Mathematically, the evolving system with homogeneous source is much simpler as one deals with ODEs, but even in that case the numerical analysis is arduous \cite{Colm12,JM98} as the non-stationary cluster mass distribution is very tricky---there is e.g. a boundary layer region describing the small mass tail of the cluster mass distribution. 

Thus we don't understand the behavior of the generalized Brownian coagulation processes with a localized source in the region $a\geq \frac{1}{2}$ which includes $a=\frac{1}{2},~\approx \frac{3}{5}, ~1$ arising in the context of polymerization. Let us also think more carefully about the region $\frac{1}{3}<a<\frac{1}{2}$ where clusters are fairly compact, $2<D_f<3$. In this region the total number of clusters remains finite. The rate equation framework is deterministic, so it is applicable only to systems with infinitely many clusters. For finite systems there are always fluctuations. If the total number of clusters diverges as $t\to\infty$, the relative magnitude of fluctuations vanish (as $1/\sqrt{\mathcal{N}}$ in most cases). For the classical Brownian coagulation fluctuations vanish, although extremely slowly, viz. as $(\ln t)^{-1/2}$; for the generalized Brownian coagulation processes with a localized source fluctuations never vanish in the region $\frac{1}{3}<a<\frac{1}{2}$.

\section{Summary}
\label{sec:conl}

Reaction-diffusion processes driven by localized input are often found in Nature and used in industry. Some of these processes which occur e.g. in electropolishing \cite{EP}, dissolution \cite{D87}, corrosion \cite{KM91}, and erosion \cite{SBG} involve a few species of atoms. These processes are well understood as they are tractable mathematically \cite{Lar,PLK:Stefan,IDLA_1,IDLA_2}. In other examples we have numerous interacting sub-species, e.g. clusters in aggregation \cite{Sid89}, aggregation-annihilation \cite{PLK:source}, and mass exchange \cite{PLK:exchange}; the analysis of these systems are much more challenging. 

Here we investigated aggregation processes driven by a localized source of monomers. When reaction rates and diffusion coefficients are mass-independent, a stationary regime is reached in the long time limit. In particular, we demonstrated that the densities of clusters of small mass exhibit an algebraic $r^{-\beta(d)}$ decay far away from the source. The decay exponent has irrational values in physically relevant dimensions: $\beta(3)=(\sqrt{17}+1)/2$ and $\beta(2)=\sqrt{8}$. The understanding of this driven aggregation process is quite complete in four dimensions where we presented the scaled mass distribution (again in the stationary regime). In three dimensions, the stationary scaled mass distribution satisfies a non-linear integro-differential equation. 

We also studied a Brownian coagulation process driven by a localized source of monomers. We argued heuristically that the stationary regime is reached, and we established chief scaling laws for a few major quantities, see Eq.~\eqref{csN}, in the classical case when clusters are spherical. If clusters are polymers undergoing Brownian coagulation, one is naturally led to the reaction rates \eqref{Brown:3d-K} and diffusion coefficients \eqref{Brown:3d-D} with $a\geq \frac{1}{2}$. In these situations the stationary regime is apparently never reached. More precisely, we provided heuristic evidence that the assumption that a stationary regime is reached leads to unphysical behaviors. Our heuristic estimates relied on scaling assumptions, so the evidence is far from strong. In particular, the marginal case of $a=\frac{1}{2}$ is known to be notoriously subtle for aggregating systems with a homogeneous source \cite{nested}, and this may be also true in our case. An obvious challenge is to explore the generalized Brownian coagulation processes in the region $a\geq \frac{1}{2}$. We also showed that even when $\frac{1}{3}<a<\frac{1}{2}$, the deterministic rate equation framework provides limited insight for the generalized Brownian coagulation process driven by a localized source. The reason is the finiteness of the total number of clusters implying the lack of self-averaging.


\begin{thebibliography}{99}
\bibitem{biol:ternary}
     Binary interactions are less common in biology. Many processes underlying cell dynamics
     are impossible without enzymes, so on the fundamental level these reaction processes 
     are at least ternary. If only two clusters undergo the final change, the coarse-grained binary
     description remains reasonable.
\bibitem{2d_exception}
     When $d=2$, the dimensions of $D$ and $K$ are the same, so the ratio $K/D$ is dimensionless and we cannot
     simultaneously set to unity both $D$ and $K$. This does not affect chief results, e.g., the density of monomers
     exhibits the $r^{-\sqrt{8}}$ decay independently on the ratio $K/D$.
\bibitem{J:eff}
     In \eqref{cr:simple}--\eqref{Nt:simple} one should actually use the renormalized source strength $J_\text{eff}$ for 
     $d>4$. It is difficult to compute $J_\text{eff}$; qualitatively, $J_\text{eff}\simeq J$ when $J\to 0$ 
     or $d\to\infty$.
\bibitem{asymp}
     The left-hand side of \eqref{CK} actually reads $\frac{d C_k}{d\ell}-\frac{1}{2}\frac{d^2 C_k}{d\ell^2}$, but
     the term with the second derivative is sub-dominant and therefore omitted.
\bibitem{van}
     It was argued \cite{vD89} that any $d>2$ dimension is critical for Brownian coagulation, so there are deviations 
     from mean-field rate equations; they are expected to be small, so we use the rate equation framework.
\bibitem{nested}
     The nested $\ln(\ln t)$ logarithmic growth was predicted \cite{JM98} for some aggregating systems 
     with homogeneous source in the marginal case.
\end{thebibliography}
\end{document}